\documentclass[aps,pra,amsfonts,amssymb,amsmath]{revtex4}

\newcommand{\be}{\begin{eqnarray}}
\newcommand{\ee}{\end{eqnarray}}

\newcommand{\n}{\nonumber}
\newcommand{\bra}[1]{\langle#1|}
\newcommand{\ket}[1]{|#1\rangle}

\usepackage{graphicx}
\usepackage{epsfig}

\begin{document}


\title{Quantum Walk  in Periodic Potential on a Line\\
    and a Model of  Interacting Opinions}


\author{Choon-Lin Ho}


\affiliation{Department of Physics, Tamkang University,
 Tamsui 25137, Taiwan, R.O.C.}




\begin{abstract}
Two subjects are discussed in this work: localisation and recurrence 
in a model of quantum walk in a periodic potential, and  a model of opinion dynamics with multiple choices of opinions.
\end{abstract}


\maketitle

\section{Introduction}

In this work I shall discuss two different subjects. One is on the phenomena of localisation and recurrence 
in a model of quantum walk in a periodic potential \cite{CH1}, and the other on a model of opinion dynamics with multiple choices of opinions \cite{CH2}.  In fact, our model of interacting opinions was inspired by our work on quantum walk.

While these two subjects seemed to be quite unrelated, they do share a common feature, namely, the external response of an individual of a system is effected according to the internal state of that individual. In the case of quantum walk, the individual is the walker at a particular site, the response is to move one step to the left or to the right, and the coin state at that site is the internal state.  For the opinion model, the individual is a member (or agent) of a social system, the response is which opinion to reveal to the public, and the internal state is the secret mind of the individual at  that moment.

\section{Localisation and recurrence of quantum walk on a periodic potential}

\subsection{Background}
 
Quantum walks are the quantum analogues of the classical random walks (for reviews and books , see e.g.: \cite{Kempe,VA,RNB,Portugal,MW}). 
It was originally proposed with the aim of finding 
quantum algorithms that are faster than classical algorithms for the same problem, because in general quantum walks diffuse faster than its classical counter parts. 
There are two distinct types of quantum walks, namely, discrete time quantum walks with a quantum coin  on the line \cite{ADZ,NV,LZG0,XZ,QX} and on graphs \cite{AAKV,WZTQW},
 and continuous time quantum walks \cite{FG,RSLLZG}.
 Some new quantum algorithms  based on quantum walks have been proposed. For instance, 
 a quantum search algorithm based on  discrete time quantum walk architecture 
 has been shown to gain an algorithmic speedup over classical algorithms \cite{SKW}, and a continuous time quantum walk was shown to be able to find
 its way across a special type of graph exponentially faster than any classical algorithms \cite{CCDFGS}.
 
 That quantum walks can escalate many classical algorithms lies in the fact that  in general quantum walks 
 diffuse faster than its classical counter parts.  
 For a process that gives a  symmetric distribution of the walker's positions, the tendency of diffusion can be 
 measured by the standard deviation of the position $\sigma (t)$  as a  function of time (step)  $t$.  For  classical random walk ,
  one has $\sigma (t)\propto \sqrt{t}$, but for a unbiased quantum walk on a line with a Hadamard coin (so-called Hadamard walk), one has $\sigma (t)\propto t$.  
  Analytical results  for quantum walk limit distributions have since been established \cite{Konno1,Konno2,Konno3}. 
 
 However, it has recently been shown that localisation of the quantum walker can occur in various situations.   Localisation means the absence of diffusion in a quantum walk. 
This issue of quantum walks has attracted much attention recently, both theoretically and experimentally.

Localisation phenomenon in a 2-dimensional quantum walk on  graphs was demonstrated numerically in \cite{TFMK}. Inspired by this work, Inui et al. \cite{IKK,IK} showed that the key factor behind this localisation phenomenon is the  the degeneracy of eigenvalues of the corresponding evolution operator.  The return probability of final-time dependent quantum walks was considered in \cite{IKMS}. 
In \cite{RMM} it was shown that a quantum walk  in a one-dimensional chain using several types of biased quantum coins, arranged in aperiodic sequences following the Fibonacci prescription,  can lead to a sub-ballistic wave-function spreading.   An approach based on P\'olya number to studying localisation in quantum walks was proposed in \cite{S1,S2,S3}. Localisation on the half-line was considered in \cite{KS}. Konno has proved mathematically that inhomogeneous discrete-time quantum walks do exhibit localisation \cite{Konno4}.
  
In \cite{SK1,SK2}  it was shown that  for a class of inhomogeneous quantum walks with multiple coins periodic in position, which is a generalisation of the model introduced in \cite{LS}, there could be localisation at the origin for certain choices of the parameters defining the model.  Furthermore, they have shown, through numerical studies, that the eigenvalue spectrum of such inhomogeneous walks could exhibit a fractal structure similar to that of the Hofstadter butterfly.   

 Changing a phase (i.e., imposing discontinuity) at a point in a discrete quantum walk has also been shown to result in certain localisation effect  \cite{Wojcik}. 
 Localisation is also observed in a quantum walk with  two coins operating at different times \cite{Machida}.
 In \cite{FP}, localisation in two-dimensional alternate quantum walks with time-perioidc coins was presented.
 The differences in limit distributions between the classical random walks and a few models of quantum walks  were presented in\cite{Shikano1}.  
 The above studies indicate that suitable modifications of the position and/or coin space could lead to a rich variety of possible 
 wave function evolutions of the quantum walker.
 For a recent review on various aspects of localisation in quantum walks, see e.g., Ref. \cite{Shikano2,VA}.
 
 Experimentally,  quantum walk revival was demonstrated in \cite{Xue} for a model with periodically changing coins (in time, or steps).  Theoretical explanation of this experimental results was recently given in \cite{CW}.
 
Here we present  numerical study of a model of quantum walk in periodic potential on the line.  We take the simple view that different potentials affect differently the way the coin state of the walker is changed.  
For simplicity and definiteness, we assume the walker's coin state is unaffected at sites without potential, and is rotated in an unbiased way according to Hadamard matrix at sites with potential. This is the simplest and most natural model of a quantum walk in a periodic potential with two coins.
 
It is found that for certain periodic potentials  the walker can be confined in the neighbourhood of the origin for sufficiently long times. Associated with such localisation effect is the recurrence of the probability of the walker returning to the neighbourhood of the origin.  

Our results show that it is possible, by controlling the periodicity of the periodic potential and the choice of coins, to control the motion of a quantum walker. 
 It would be nice if the quantum walks on periodic potential considered here could be experimentally implemented, say in optical lattice.
 
 A different model of quantum walk in a periodic potential on a line has also been considered in \cite{LZG}.  This model uses two different kinds of quantum walks at site with without potentials. 
 Localisation phenomenon was not reported in this work.
 
 \subsection{The Model}
 
 Our model is defined as follows \cite{CH1}.
The total Hilbert space is given
by $\mathcal{H}\equiv\mathcal{H}_{P}\otimes\mathcal{H}_{C}$, where
$\mathcal{H}_{P}$ is spanned by the orthonormal vectors $\{\ket{x}, x=0,\pm 1, \pm 2, \ldots\} $
representing the positions of the walker, and $\mathcal{H}_{C}$
is a two-dimensional coin space spanned by two orthonormal vectors
denoted by $\ket{0}$ and $\ket{1}$. The dynamics of the walk is controlled by a coin flip operator $C$, which modifies the coin states of the walker, and a
conditional shift operator $S$ that shifts the walker's position according to the latest state of the coin.  

Thus the evaluation operator for one step of walk is
$U=S\cdot (C\otimes I)$.   If the initial state of the walker and the coin  is $\ket{\psi_0}$,  then after $t$ steps of the walk the state of the system is 
$\ket{\psi (t)}=U^t\ket{\psi_0}$.

To  define a model of quantum walk in  a periodic potential on a line, we take the simple view that different potentials affect differently the way the coin state of the walker is changed.  Thus we suppose the coin state is changed by a coin operator $C_0$ when there is no potential,  and by $C_p$ when the field is present.  The 
position displacement operator at each position is
\begin{equation}
S_x=\ket{0}\bra{0}\otimes \ket{x+1}\bra{x} + \ket{1}\bra{1}\otimes \ket{x-1}\bra{x}.\n
\end{equation}
Together the evolution operator is
\be
U=\sum_{x: {\rm no\  potential} }S_x\left(C_0\otimes I \right)  + \sum_{x: {\rm at\  potential} }S_x\left(C_p\otimes I \right). 
\n
%
\ee

In this work, for simplicity and definiteness, we shall choose $C_0=I$ and $C_p=H$, where $H$ is  the Hadamard matrix
\begin{eqnarray}
H=
 \frac{1}{\sqrt{2}}
 \left( \begin{array}{cc}
  1 & 1 \\
  1 & -1
  \end{array}\right).
  \n
\end{eqnarray}
This means that the walker's coin state is unaffected at sites without potential, and is rotated in an unbiased way according to the
Hadamard matrix at sites with potential. 

This is the simplest and most natural model of a quantum walk in a periodic potential with two coins.

The state of the walker after $t$ steps is
\be
\ket{\psi(t)}=U^t \ket{\psi_0}
=\sum_{x=-\infty}^\infty \left[A_x(t) \ket{0} + B_x(t) \ket{1}\right]\,\ket{x},\n
\ee
where $\ket{\psi_0}$ is the initial state, and
\be
\sum_x |A_x(t)|^2 + |B_x(t)|^2 =1.\n
\ee
At time $t$ the probability that the walker is to be found at position $\ket{x}$ is
\be
 p(x,t)= |A_x(t)|^2 + |B_x(t)|^2.\n
\ee

\subsection{Numerical results}

We shall be interested in symmetric walks such that $\langle x(t) \rangle=0$.  
It is known that  the  initial state of the walker and coin given by
\begin{equation}
\ket{\psi_0}=\frac{1}{\sqrt{2}}\left(\ket{0} + i \ket{1}\right)\ket{0}\n
\end{equation}
gives rise to an outgoing symmetric probability distribution
on the positions when a single Hadamard coin is used.
So the main physical quantity to characterise the quantum walk is the standard 
deviation $\sigma(t)=\sqrt{\langle x(t)^2 \rangle}$.  Also, as we have in mind the possibility of localisation of the walker at the origin, we shall also consider the probability $P_0(t)$ of the walker at the origin $x=0$ as a function of the step $t$.

 We have considered six generic cases of quantum walk in periodic potential on a line with a period $N$.  
For simplicity, we shall adopt the notation $[C_1:q, C_2: (N-q)]$ to denote the situation where the coin operator $C_1$ is to be used in the first $q$ positions  and the coin $C_2$ is used in the remaining $N-q$ positions.  The origin $x=0$ is always assumed to be at the middle-point of the first $q$ positions (so in this work $q$ is always taken to be odd). 

The six cases are:
\be
{\rm \  IA}:   [H:1,~~I: N-1], ~ &N\geq 2 ;\n\\
{\rm \  IB}:   [I:1,~~H: N-1], ~  &N\geq 2 ;\n\\
{\rm \  IIA}:   [H:N-1,~~I: 1], &N={\rm even} ;\n\\
{\rm \  IIB}:   [I:N-1,~~H: 1],  & N={\rm even}; \n\\
{\rm \  IIIA}:   [H: q,~~I: q],~~~~  & q={\rm odd}; \n\\
{\rm \  IIIB}:   [I: q,~~H: q],~~~~  &  q={\rm odd}; \n
\ee


In Fig.\,1 we show the standard deviation $\sigma (t)$ versus the number of step $t$.  It is seen that, as in the standard unbiased 
Hadamard quantum walk, $\sigma (t)$ is generally asymptotically linear in $t$, i.e. $\sigma(t)\propto t$. 
 
Note that the graphs of $\sigma (t)$ for quantum walks of case IA and IIB overlap with that for the Hadamard walk at large $t$. 
For the other cases, the slope of the $\sigma (t)$ are smaller.   This means that diffusion in cases other than IA and IIB are slower, with case IIIB being the slowest.  Nevertheless, standard deviations of diffusions in all these cases are still larger than that in the classical walk given by $\sigma(t)=\sqrt{t}$).

\begin{figure}[hbt]
 \centering
 \includegraphics*[width=14cm,height=15cm]{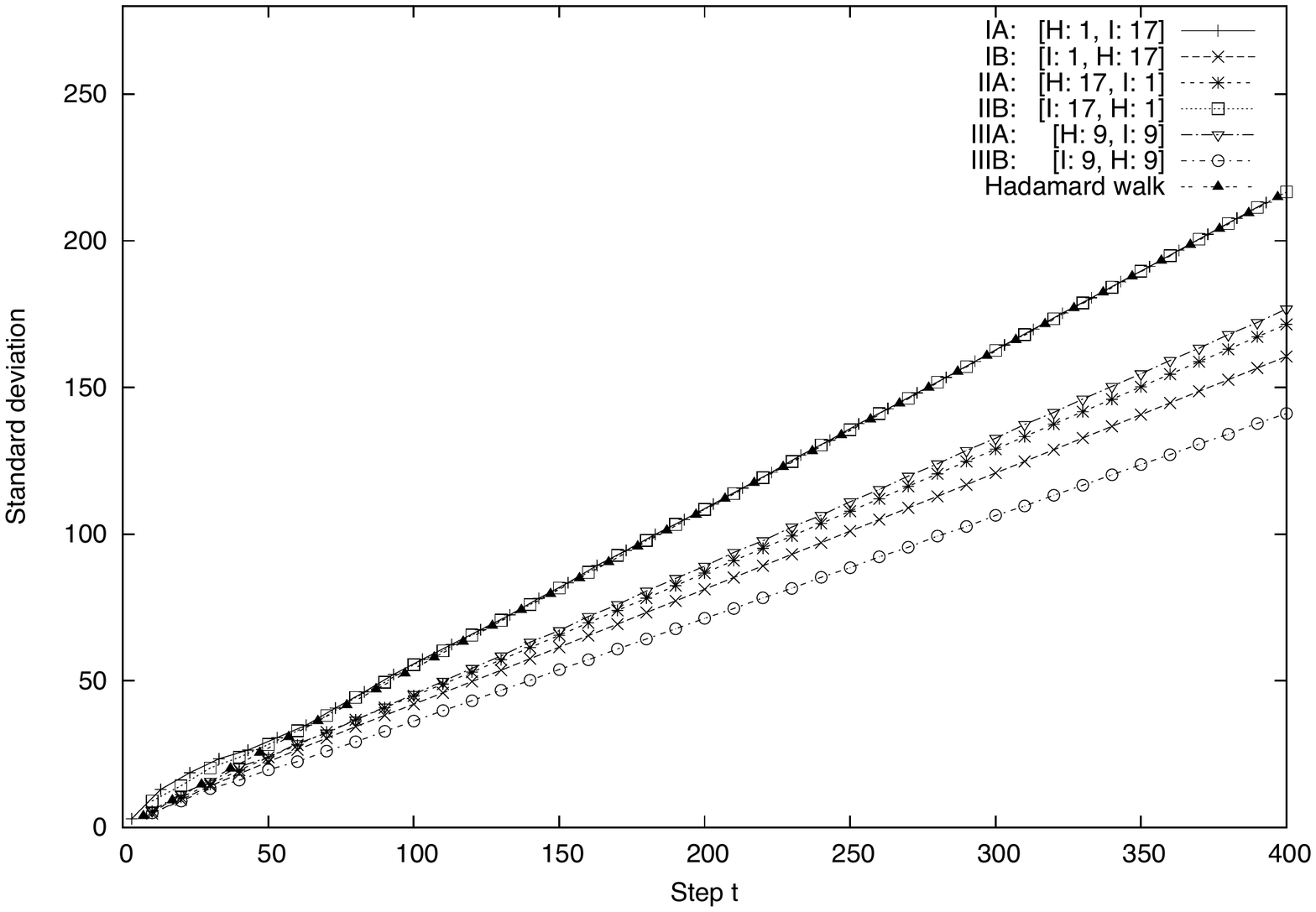}\\
{Fig.\,1: The standard deviation $\sigma (t)$ versus the number of step $t$.}
\end{figure}

%
%


\centerline{\bf {Cases IA and IIIB}}

~~~~ In Figs.~2 to 4 below we present some figures to contrast the behaviours of two extreme cases, i.e., Cases IA and IIIB, which have the the largest and smallest $\sigma(t)$, respectively. 

\begin{enumerate}

\item[$\bullet$]

For case IA, the walker encounters potential field only at the positions $\ket{x}$ such that $x$ is a multiple of the period $N$.
It is found that two peaks symmetric with respect to the origin just move away from the origin as in the case of the Hadamard walk.

\item[$\bullet$]

  For case IIIB, the  periodic potential (period $N=2q$) where locations with and without potential have equal length  $q$, and  the walker begins its walk at the middle of zero  potential region.   Localisation and recurrence occur  more  significantly for case IIIB. In fact, localisation occurs already for $q>3$.  Thus the tendency of diffusion is much slower in this case than in other cases.  This is also reflected in the fact, as depicted in Fig.~1, that  the standard deviation $\sigma (t)$ of this case has the smallest slope than those of the other cases.

\end{enumerate}

\begin{figure}[ht]
 \centering
\includegraphics*[width=8cm,height=10cm]{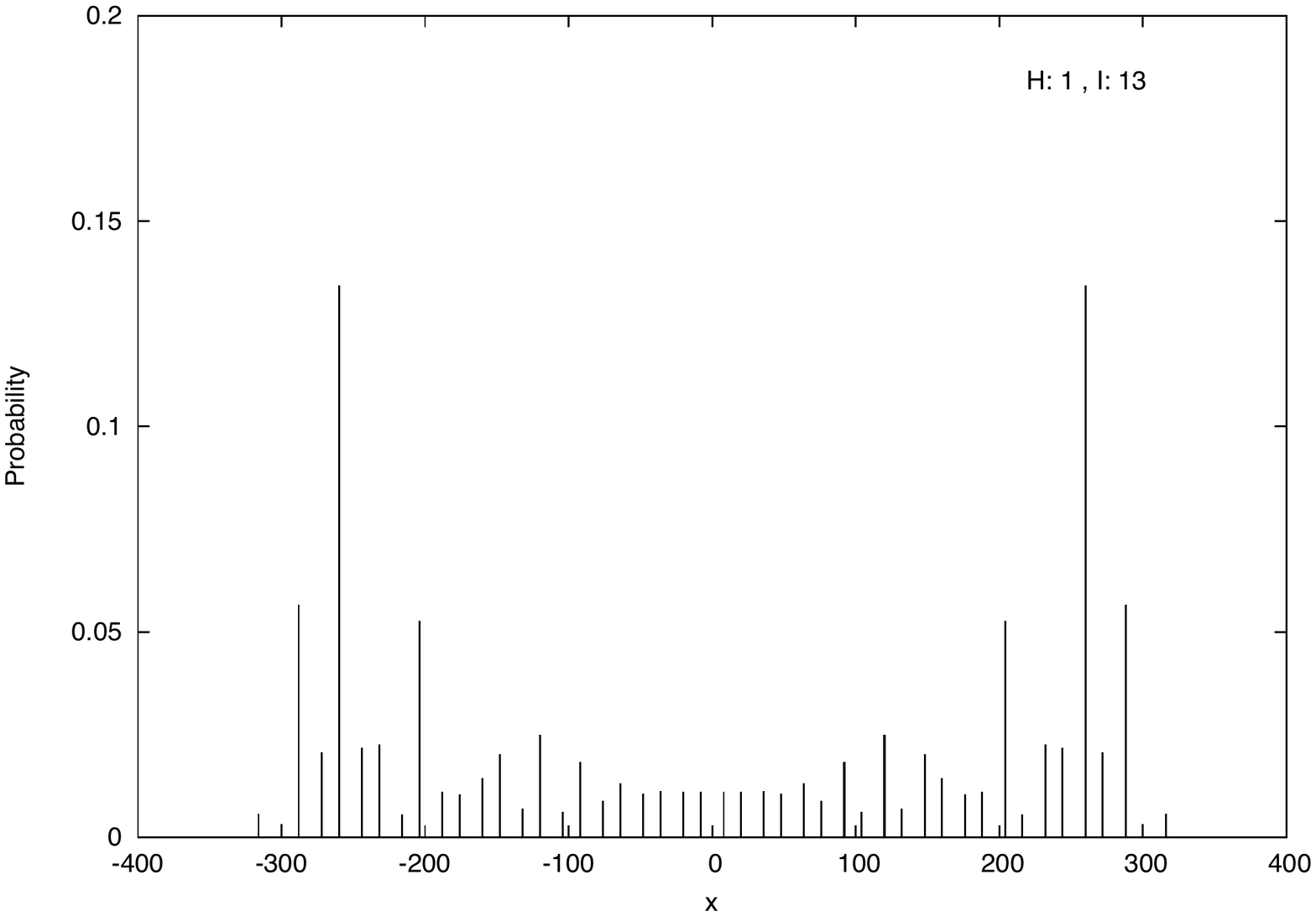}\hspace{0.5cm}
\includegraphics*[width=8cm,height=10cm]{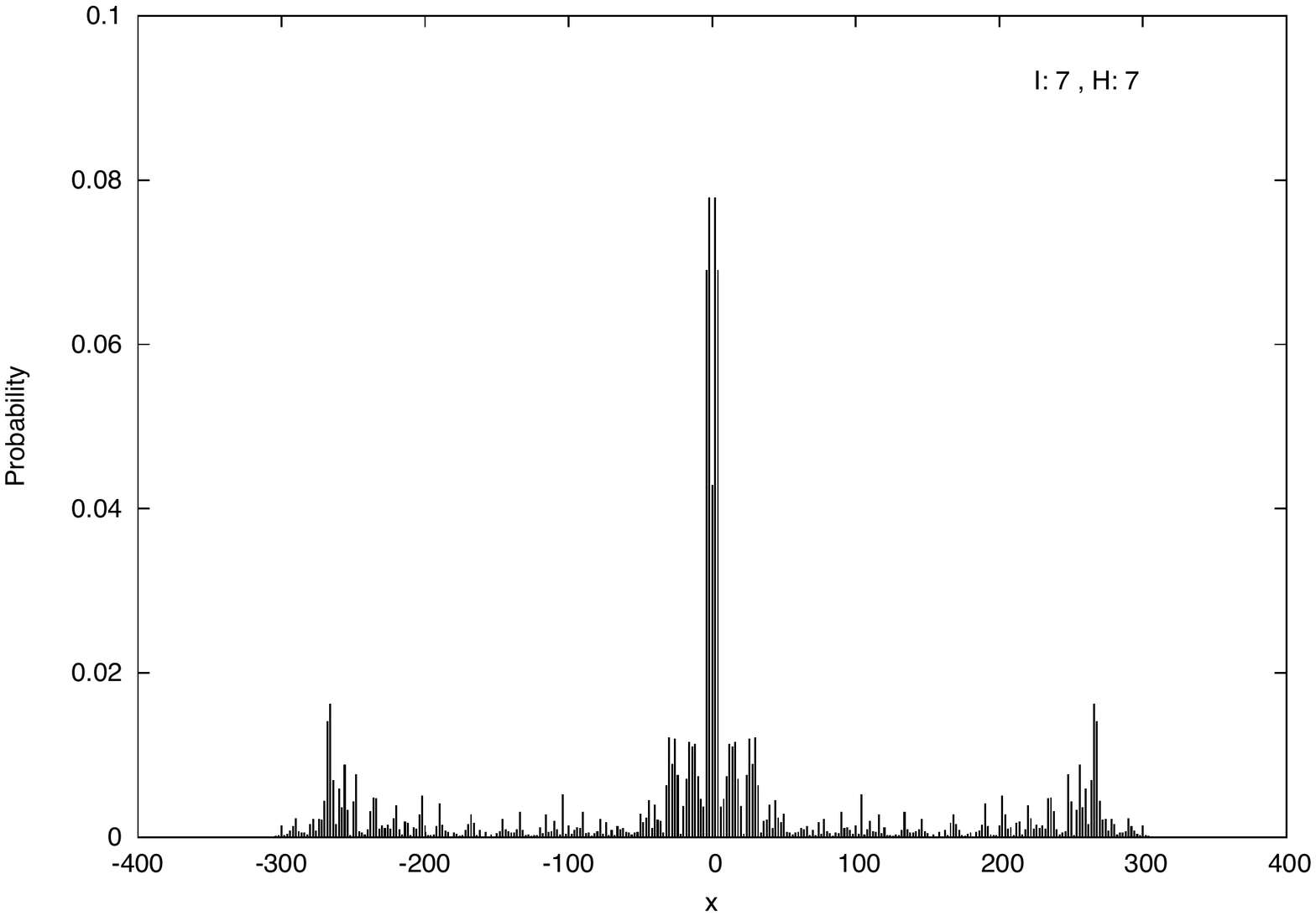}\\
{Fig.\,2:  Probability $P(x,t)$ of the quantum walk of case IA (left) and case IIIB (right) with period $N=14$ at  $t=400$. Note that the probability at positions with odd $x$ are zero. There is no localisation for case IA, while localisation at the origin is rather significant for case IIIB when $q=N/2>3$.}
\end{figure}

\begin{figure}[ht]
 \centering
 \includegraphics*[width=14cm,height=15cm]{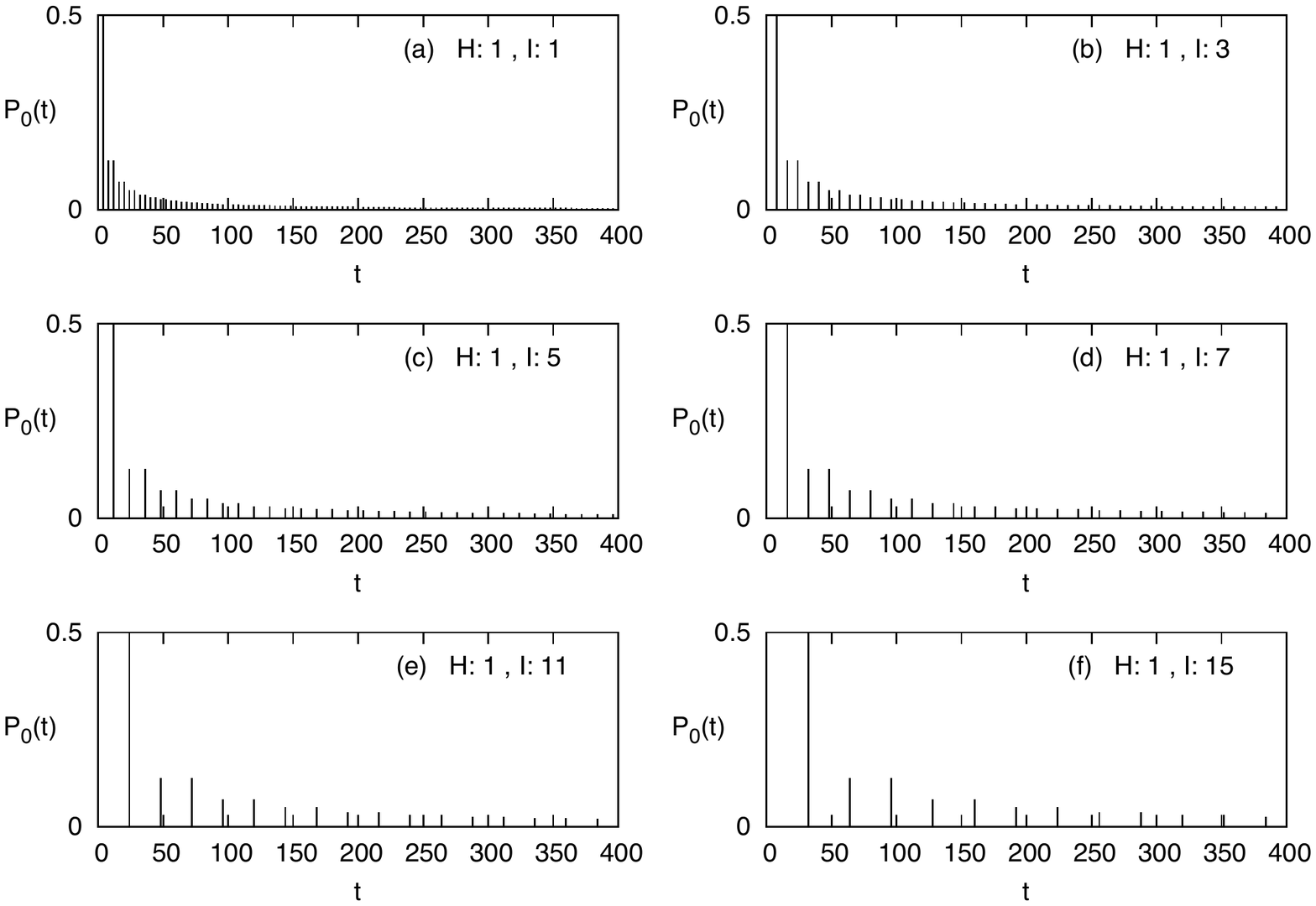}\\
{Fig.\,3a: Probability $P_0(t)$ at the origin of the quantum walk of  case IA.
The number of steps is taken up to 400.  Note that the probability at the origin must be zero when the step $t$ is odd. This is the same for the all other cases considered in this work. There is no localization as $P_0(t)$ dies away at large steps.}
\end{figure}

\begin{figure}[ht]
 \centering
 \includegraphics*[width=14cm,height=15cm]{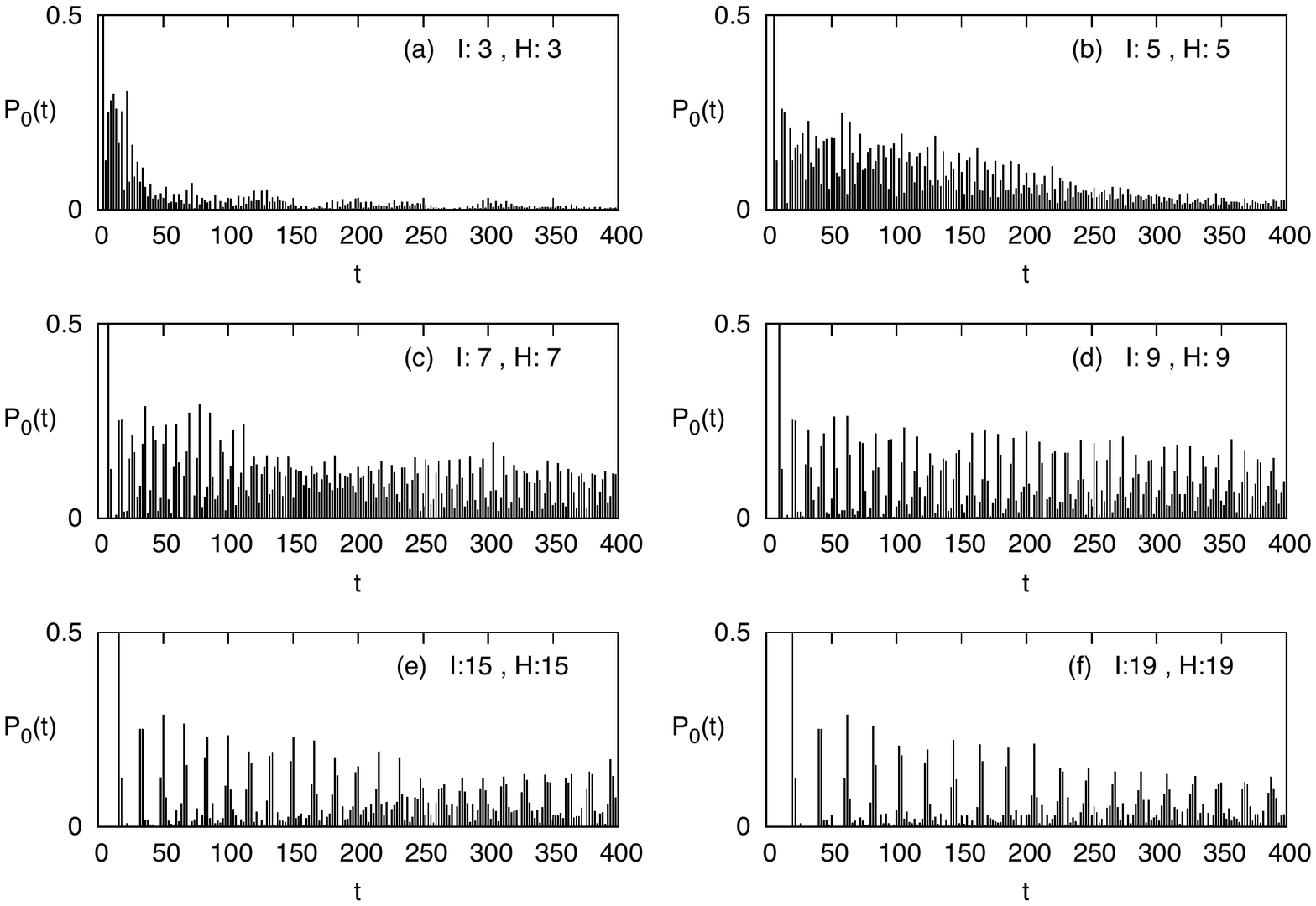}\\
{Fig.\,3b:  Probability $P_0(t)$ at the origin of the quantum walk of  case IIIB.
The number of steps is taken up to 400. Significant recurrence occurs for $q>3$.}
\end{figure}



\subsection{Discussion}

Six generic cases of such quantum walks were studied numerically in our work.  It is found that of the six cases, two cases, case IA and IIB, behave in
a similar pattern as the original Hadamard walk where only a Hadamard coin is being used throughout the walk on the line. They show the 
same asymptotic values of the standard deviation as function of the step.  On the other hand, in the other four cases,   localisation effect is possible, where the walker could be confined in the neighbourhood of the origin for sufficiently long times. Associated with such localisation effect is the recurrence of the probability of the walker returning to the neighbourhood of the origin. 

A notable difference between the  cases IA and IIB with the other four cases is the number of the positions with potential, or the points where Hadamard coin is used within a single period of the potential:  it is smaller than the number of points without potential (or points using identity coin).  Also, in the other four cases, localisation and recurrence occur only when the number of points with Hadamard coin is ``sufficiently" larger than the number of points using identity coin.  This implies the existence of critical values of period $N$ for these cases.

Furthermore, it is also observed that the effect of  localisation and recurrence occur more strongly in case IIIB and IB, as shown in Fig.\,1. In these two cases, the walker starts at the centre of the valley (with identity coin) of the potential. 

To summarise, it appears that in the model of quantum walk in periodic potential proposed in this work, localisation and recurrence effect are stronger if the walker begin its walk in the middle of the valley (with identity coin) of a periodic  potential with a larger portion of potential (with Hadamard coin).

It is worthwhile to note that the phenomena of recurrence and localisation in the quantum walks reported here refers to the repetitive pattern of high probability of finding the quantum walker being found at its initial position. This is in great contrast to the symmetric moving away pattern of the Hadamard quantum walk, and is closer to the classical random walk in which the walker has a high probability (the Gaussian distribution) to return to the origin. But just like the random walk, the recurrent quantum walks are still diffusing.
Fig. 1 showed that all the six cases studied here diffuse faster than the classical random walk.

One may wonder if the recurrence reported here is related to the well-known phenomenon of collapses and revivals of under intense study in quantum optics.  The latter phenomenon is mainly associated with the approximate  or exact return to its initial wave form of a wave-packet during its evolution , especially the evolution of wave-packet formed from a superposition of states heavily peaked around an eigenstate \cite{Eberly}.  Collapses and revivals in these cases are simply due to the interference effect of the relative phases of the eigenstates in the wave-packets.  As mentioned in the last paragraph,  in the quantum walk model reported here, recurrence and localisation refer to the repetitive pattern of high probability of finding the quantum walker being found at its initial position (the origin in this work).  We believe that the recurrence and localisation of the quantum walks reported here is due mainly to the interference of the left-moving and right -moving waves with amplitude and phases modified  by the coin operator at each lattice site. 

As a first attempt to study localisation and recurrence phenomena of quantum walks in periodic potentials, we have relied on numerical method so far.  It is desirable to be able to study this problem  with an analytic approach, in order to have a fundamental understanding of the phenomena of localisation and recurrence in 
space-inhomogeneous quantum walks of which our model is only an example.  Very recently, periodicity for the Hadamard walks on cycles has  been studied analytically in \cite{Konno5}. It would be nice to see if the approach in this work could be extended to 
periodicity of the Hadamard walks on a line as presented here.

 It is also interesting to study the behaviours of the quantum walk with other choices of the coins $C_0$ and $C_p$.   Generalisation of the  present work to higher-dimensional cases is straightforward.   It would be nice if the quantum walks on periodic potential considered here could be experimentally implemented, say in optical lattice.

\section{A model of Interacting opinions}

\subsection{Background}

 In many daily situations it is necessary for a group of people to reach shared decisions or opinions. More often than not, people have different opinions. For instance, whom to vote for before an election; which restaurant to go for dinner, etc.
Despite this,  quite often spontaneous agreement (consensus/majority opinion) can later be reached through discussions/debates etc.
     
The emerging subject of opinion dynamics is concerned with setting up dynamical models to understand how such consensus emerges, or opinion formation. 
Here the society is regarded as a system in which consensus may result from repeated local interactions among the individuals (the agents).
Opinion in the mind of any individual has to be quantified by a variable $c$, say, in order for any model building.  Depending on whether $c$ is discrete or continuous, one has model of discrete or continuous opinions formation. Discrete opinions can be binary, with $c=0,1$, representing situations which call for decisions such as yes/no, believe/disbelieve, vote for/against, etc.  It can also be multiple-valued ($c=1,2, \ldots, q$), representing situations in which the degree of preference of an opinion can have $q$ different levels.  The continuum limits of the latter
lead to models with continuous opinions, say with $0\leq c \leq 1$ after appropriate scaling.  This is the case, for instance, when one is asked to rate a book or a movie, ... etc.

 In the past years physicists have attempted to use statistical physics as a framework to study collective phenomena emerging from the interactions of individuals as elementary units in social structures (for a review on the subject, see e.g., \cite{CFL,SC}). A general discussion on collective phenomena in social psychology can be found in \cite{G1}.

One of  the oldest models of discrete opinions which mimics binary opinions systems is the voter model \cite{voter} that represents a society in which each agent always follows the opinion of one of his/her nearest neighbours.  The Sznajd model describes a society in which agents are influenced not by any individuals but by groups \cite{Sznajd}.  The majority-rule model mimics a society in which the majority opinion of a randomly chosen group within the total population is assigned to all the agents of that group \cite{G2,G3,G4}.  The Society impact theory describes how individual feel the presence of their peer and how they in turn influence other individuals \cite{SIT}.

The most well-known models of  continuous opinions are the so-called Bounded confidence models. These models consider the realistic aspect of human communication that, while any agent can talk to every other agent, a real discussion  exists only if the opinions of the people involved are sufficiently close to each other.  The two most popular such models are the Deffuant model \cite{D} and the Hegselmann-Krause model \cite{HK}.
In the Deffuant model an agent interacts  only with his/her nearest neighbours, whilst in the Hegselmann-Krause model an agent simultaneously interacts with all other agents whose opinions are within certain prescribed bound. 
More recently, a random  kinetic-exchange type model of continuous opinions is proposed in \cite{LCCC}. Multi-dimensional, or vectorial extensions of the bounded confidence models have also been considered in \cite{FLPR,L}.

As far as we know, all the models proposed so far were for a single choice of opinion, be it discrete (binary or multuple-valued) or continuous. However, in real life, there are situations in which one is faced with multiple choices of opinions. For instance, whether to eat (A) an apple or (B) an orange; to make a choice between (A) hiking in the mountains or (B) having a picnic by the seaside. One can assign binary or continuous values to both opinions. 

We have proposed a model of interacting multiple choices of opinions. As the discrete case can be considered a special case of the continuous one, we shall present the model only for the case of multiple choices of continuous opinions.  We have in mind the situation best illustrated as follows.

Consider a group of kindergarten children lining up in a circle. They are given two choices to decide for the day's activity: Choice (A): hiking in the mountains, and Choice (B): picnic by the seaside. At each step everyone of them will reveal his/her most preferred choice by showing a card, say ``Red" card for the choice A and ``Blue" card for the choice (B).  So each child knows the majority preference at any time.  This awareness of others' choices is the so-called social pressure.  The process of opinion propagation starts by randomly choosing a child as the first lobbyist to persuade his/her nearest neighbour, the listener. If the listener is convinced, then he/she will enhance the weight of the choice preferred by the lobbyist, and reduce the weight of the other choice according to certain rule. He/she then updates the colour of the his/her card, and proceed to persuade his/her nearest neighbour (in the original direction of opinion propagation). If the listener is not convinced, then he/she will retain the colour card, and try to persuade the lobbyist to adopt his/her preferred choice.  In this latter case, the direction of persuasion is reversed.  And this process goes on until a stationary state is attained. 
The stationary state may be one that consists of divided opinions among the agents, or one with consensus on one of the two choices.


%
%

\subsection{The model}

For simplicity and definiteness, we shall discuss the situation where only two different opinions, $A$ and $B$, need be decided. Generalisation to more opinions is straightforward.

To impose a one dimensional periodic boundary condition, we assume that $N$ agents are arranged in a circle.
At each step of the process every agent has in mind certain degree of preference of the two opinions $A$ and $B$. 
Every agent will reveal his/her preferred choice according to which opinion has the greater weight at that moment. The ratio of the two revealed opinions serves as a reference to every agent in the next step in his/her decision as to whether he/she want to change opinion when persuaded by his/her neighbour. 

Our model is described below in three steps: 1) set up of  the system, 2) persuasion process, 
and 3) internal transformation of opinion state and propagation of opinion.

\subsubsection{System set up}

\begin{enumerate}

\item[$\bullet$]

{\bf Representation of opinion state}:

 ~~~The opinion in the mind of any agent $(i=1,\ldots, N)$ at each step $t=0,1,2,\ldots$ can be represented by a two-component state function $\psi_i (t)$ as
\be
 \psi_i (t) =
 \left( \begin{array}{c}
 c_{i,A}(t) \\
 c_{i,B} (t) \end{array}\right).
\end{eqnarray}
Here $ 0<c_{i,A}(t), c_{i,B}(t) <1$ give the measure of the degree of preference  of  $A$ and $ B$ of the $i$-agent at step $t$.


\item[$\bullet$]

{\bf Revealed preference}:

~~~At each step every agent will show his/her preference to the other agents. Which choice of opinion $A$ and $B$  is shown to the others 
is determined by the factor
\be
p_{i, \tau}(t)\equiv \frac{c_{i,\tau}(t)}{c_{i, A}(t)+c_{i,B}(t)},~~\tau=A, B.
\ee
When  $ p_{i, A}(t) > p_{i, B}(t)$, the agent will show the choice $A$, otherwise he/she will show $B$ to the others. 

~~~The revealed preference of opinion $A$ is defined by the ratio of the total number of agents choosing $A$ at step $t$  among the $N$ agents:
\be
p_{s, A}(t)\equiv \sum_{i=1}^N \frac{[p_{i,A}(t)+0.5]}{N},
\ee
where $[x]$ represents the integral part of $x$. 
The corresponding revealed preference of opinion $B$ is simply given by $ p_{s, B}(t)=1-p_{s, A}(t)$. 
These revealed preferences in some sense represent the social pressure to each and every agent in their decision in the next step.

~~~When either $p_{s, A}(t)=1$ or $p_{s, B}(t)=1$, the group of agents reaches a consensus of opinion $A$ or $B$, respectively.

\end{enumerate}

\subsubsection{Persuasion process}
 
\begin{enumerate} 
 
\item[$\bullet$] {\bf Initial setting}:
 
~~~The initial opinion states of every agent, characterised by the coefficients $c_{i,A}$ and $c_{i,B}$,  will be randomly generated. 
One then randomly selects an agent $j$ and his/her neighbour $i$($i=j+1\: {\rm or}  \:j-1$) as the first lobbyist and the first listener, respectively.

\item[$\bullet$] {\bf Decision factor}:

~~~We shall consider in this paper the situation where the listener is more likely to change his/her mind if his/her preference differs too much from the majority opinion revealed (social pressure/bandwagon effect), and is more likely to be convinced by the lobbyist if their opinions do not differ too much (peer effect). 

~~~The influence of these two factors on the opinion state of the listener can be quantified by a decision factor, $P_{i,\alpha}(t)$, representing the social pressure (gauged by a parameter $0\leq \alpha\leq 1$) and 
the influence of the lobbyist ($j$-agent) on the $i$-person:
\be
P_{i,\alpha}(t)\equiv \alpha |p_{s,A}(t)-p_{i,A(t)}| + \left(1-\alpha\right) \left(1-|p_{j,A}(t)-p_{i,A}(t)|\right).
\label{factor}
\ee
The first term represents the difference between the opinions of the agent and the majority of the group, and the second term gauges the difference between the opinions of  the agent and the persuader. 
$P_{i,\alpha}(t)$ is large when the preference of the agent differs greatly from that of the majority and closes to that of the lobbyist. 

~~~We then compare $P_{i,\alpha}(t)$ with a reference (random) number $ r(t)\in [0,1]$ (accounting for the state of mind/mood of the listener at that time), 
\be
&  P_ {i,\alpha}(t) >r (t) &: {\rm\ persuasion\  successful};\n \\
& P_{i,\alpha}(t)< r(t) & : {\rm\ persuasion\  failed}.
\label{compare}
\ee

\end{enumerate}


\subsubsection{Transformation of opinion states and Propagation of opinion}

Depending on whether the listener is convinced or not by the lobbyist, he/she will have to modify his/her opinion state and decide to whom he/she should  persuade.   For our model we shall adopt the following rules:

\begin{enumerate}
\item[$\bullet$] 
 $P_{i,\alpha}(t)>r (t)$ : persuasion successful, the $i$-agent will update  his/her opinion state, and proceed to persuade his/her neighbour in the same direction, i.e., the $i+1$-agent away from the original lobbyist.

~~~For simplicity, we assume in this model the opinion state of the $i$-agent is updated according to the influence of the lobbyist, proportional to the  difference in their opinions measured by $( p_{j,A}(t)-p_{i,A}(t))$ and gauged by a parameter $0\leq \mu\leq 1$. :
\be
c_{i,A}(t+1)&=&\left[ c_{i,A(t)}+\mu \left( p_{j,A}(t)-p_{i,A}(t) \right)\right],~~([.]: {\rm integral\  part})\n\\
 c_{i,B}(t+1) &=&\left[ c_{i,B}(t)+\mu \left(p_{j,B}(t)-p_{i,B}(t) \right)\right]\n\\
                      &=&\left[ c_{i,B}(t)-\mu \left( p_{j,A}(t)-p_{i,A}(t) \right)\right],~({\rm note}\  p_B(t)=1-p_A(t))\n
\ee

Taking the integral part is to ensure that the coefficients stay in their defined range, i.e., when the coefficient $c_{i,A}(t)$ (or $c_{i,B}(t)$) $ >1$ (or $<0$), it is set to
$ 1 (0)$.

\item[$\bullet$]  $P_{i,\alpha}(t) <r(t)$ : persuasion failed, the $i$-agent retains opinion state, and proceeds to persuade neighbour in the reversed direction, i.e., the original lobbyist.

 \end{enumerate}
 
Stationary state is reached when the opinion states of these $N$ agents become fixed.
Consensus is defined as the state when all the $N$ agents agree on the same opinion, either $A$ or $B$.

%
%

\subsection{Numerical results}

We have simulated the model described above for a number of parameters $\alpha, \mu$ and $N$. The main properties under study are whether stationary state, and state with consensus in particular, can be attained, and the average time (number of steps) to attain it. For each set of parameters, a large number of initial opinions is randomly set up, and the ensemble average of the time needed to reach stationary state/consensus state is computed.

In Fig.\,4  we show the time, or rather the number of steps, required for a system of $N=25$ agents to reach a consensus with different choices of $\alpha$ and $\mu$.
The figure on the left  shows the situations in which the initial majority opinion, here the choice B, is enhanced and finally became the final consensus, while the one on the right shows that there are situations where consensuses were  reached only after several exchanges of majority opinions,. The final consensus could be a different choice with the same set of parameters.

In Fig.\,5 we show the pictures of  frequencies of the values $(c_A, c_B)$ for $\alpha=0.5$, $\mu=0.5$. This gives an idea of how these coefficients evolve. Consensus is reached when all the points are located below (with choice A) or above (with choice B) the line $c_A=c_B$.  Note that in our model consensus is defined by all $c_A>c_B$ or $c_B>c_A$.  Hence it is not necessary, as is evident from the figures, that $c_A=1$ or $c_B=1$ for the all agents in the final state.

Lastly, in Figs.\,6 we present 3-dimensional plots of $\langle T\rangle$ versus $\alpha$ and $\mu$ with an ensemble size of $200$ for each set of parameters.  It is seen that   
$\langle T\rangle$ is higher near $(\alpha, \mu)=(0,0)$ and $(1,0)$.  This is understood as the listener being harder to persuade when he/she is indifferent to the lobbyist's opinion ($\mu\sim 0$), or  he/she is very concerned with the public opinion.  The average consensus time needed also increases as the number of agent $N$ increases.  This is in conformity with the usual experience.

\begin{figure}[ht] \centering
\includegraphics*[width=8cm,height=8cm]{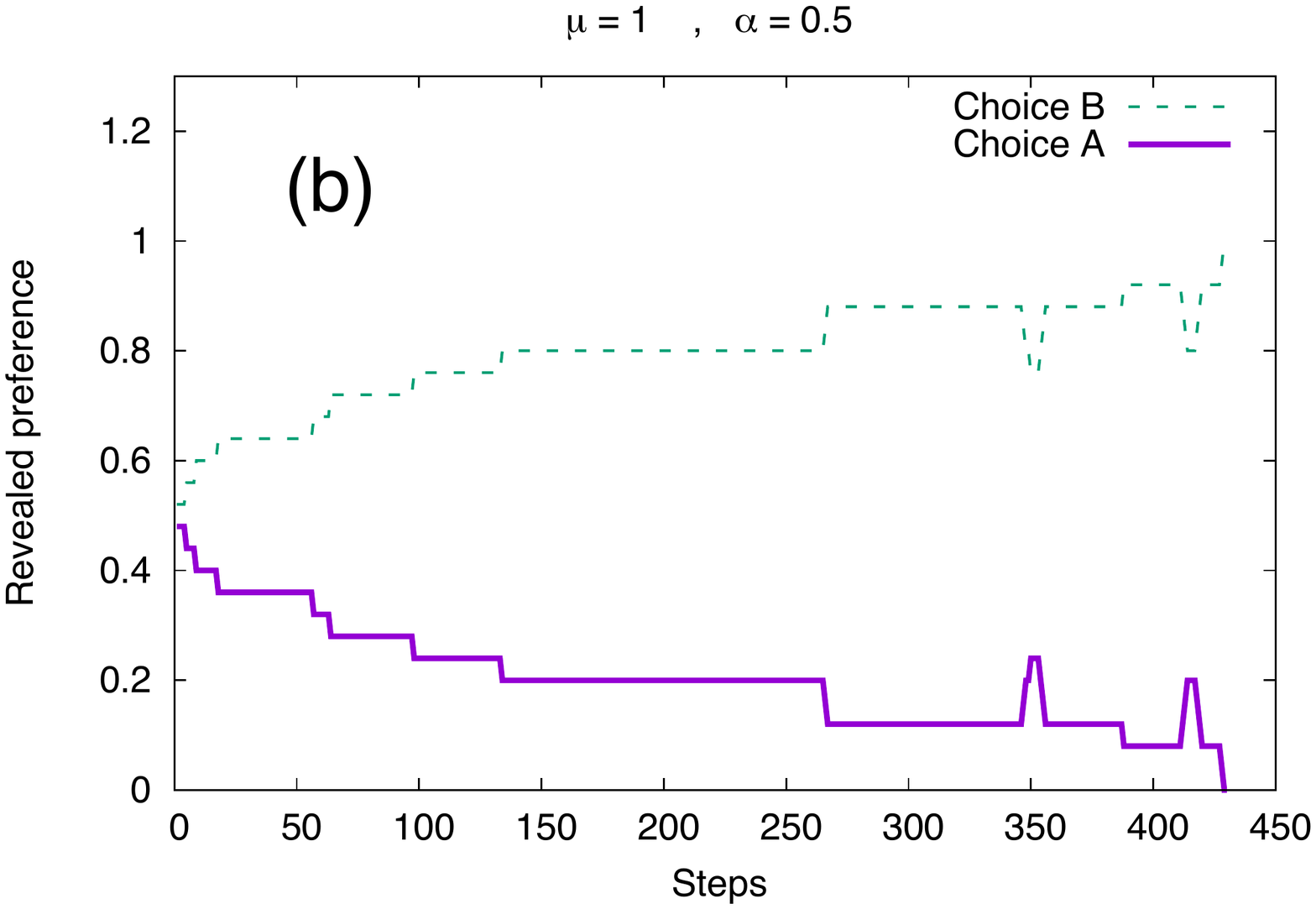}\hspace{0.5cm}
\includegraphics*[width=8cm,height=8cm]{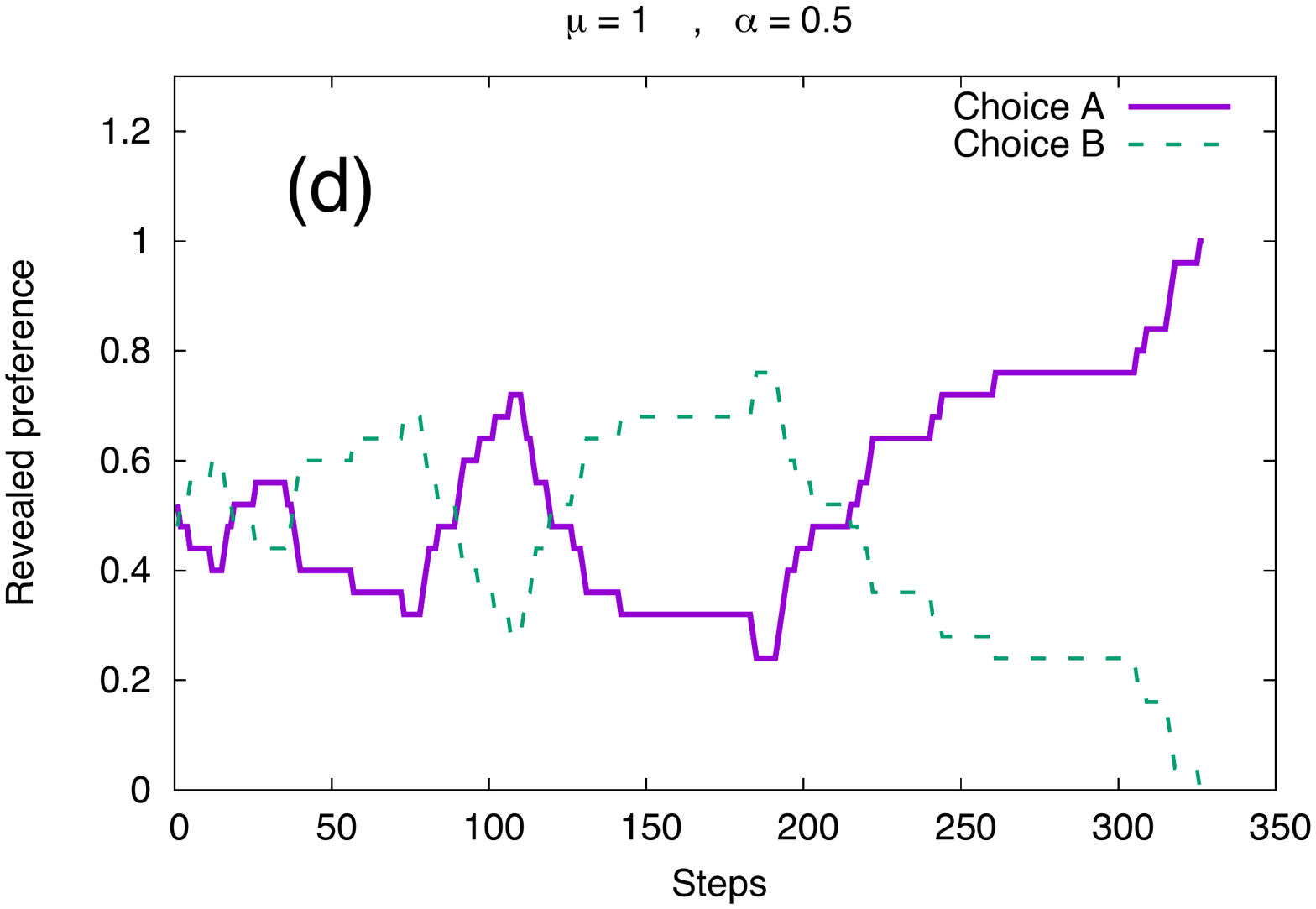}\\
{Fig.\,4: Plot of revealed preference $c_{A,B}$ versus number of steps $T$ 
with $\alpha=0.5$, $\mu=1$.} 
\end{figure}

\begin{figure}[ht] \centering
\includegraphics*[width=8cm,height=8cm]{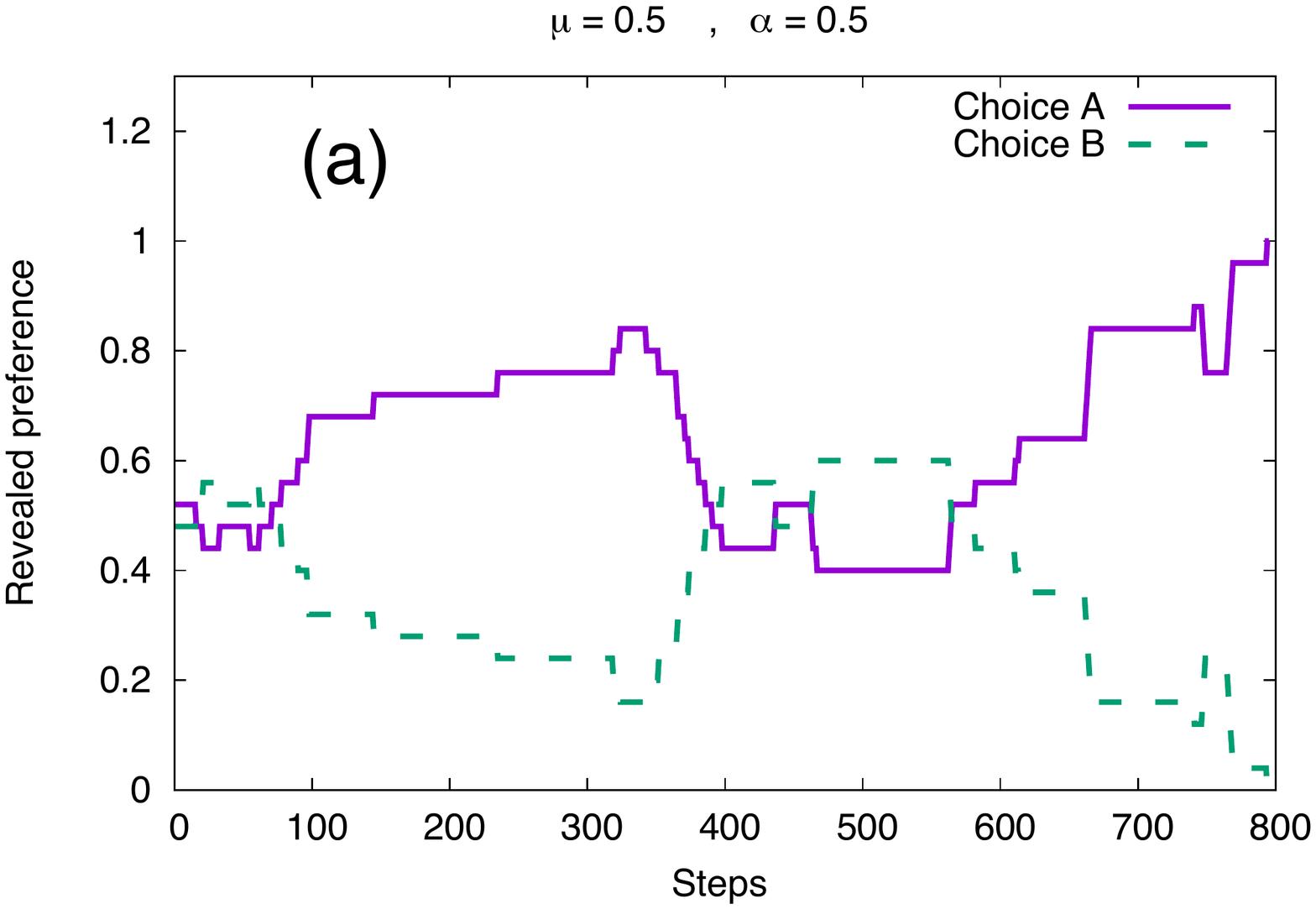}\hspace{0.5cm}
\includegraphics*[width=8cm,height=10cm]{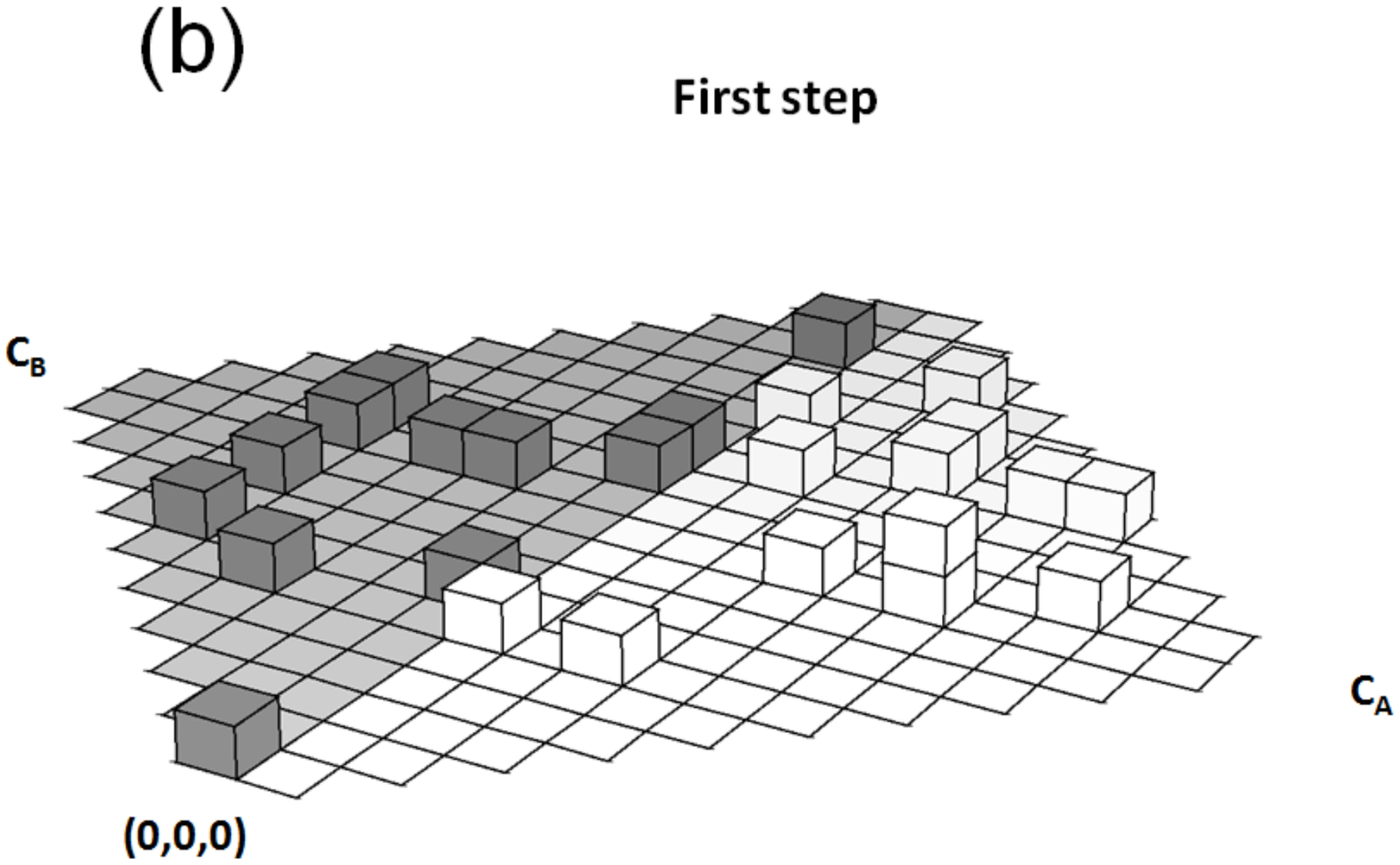}\\
\includegraphics*[width=8cm,height=8cm]{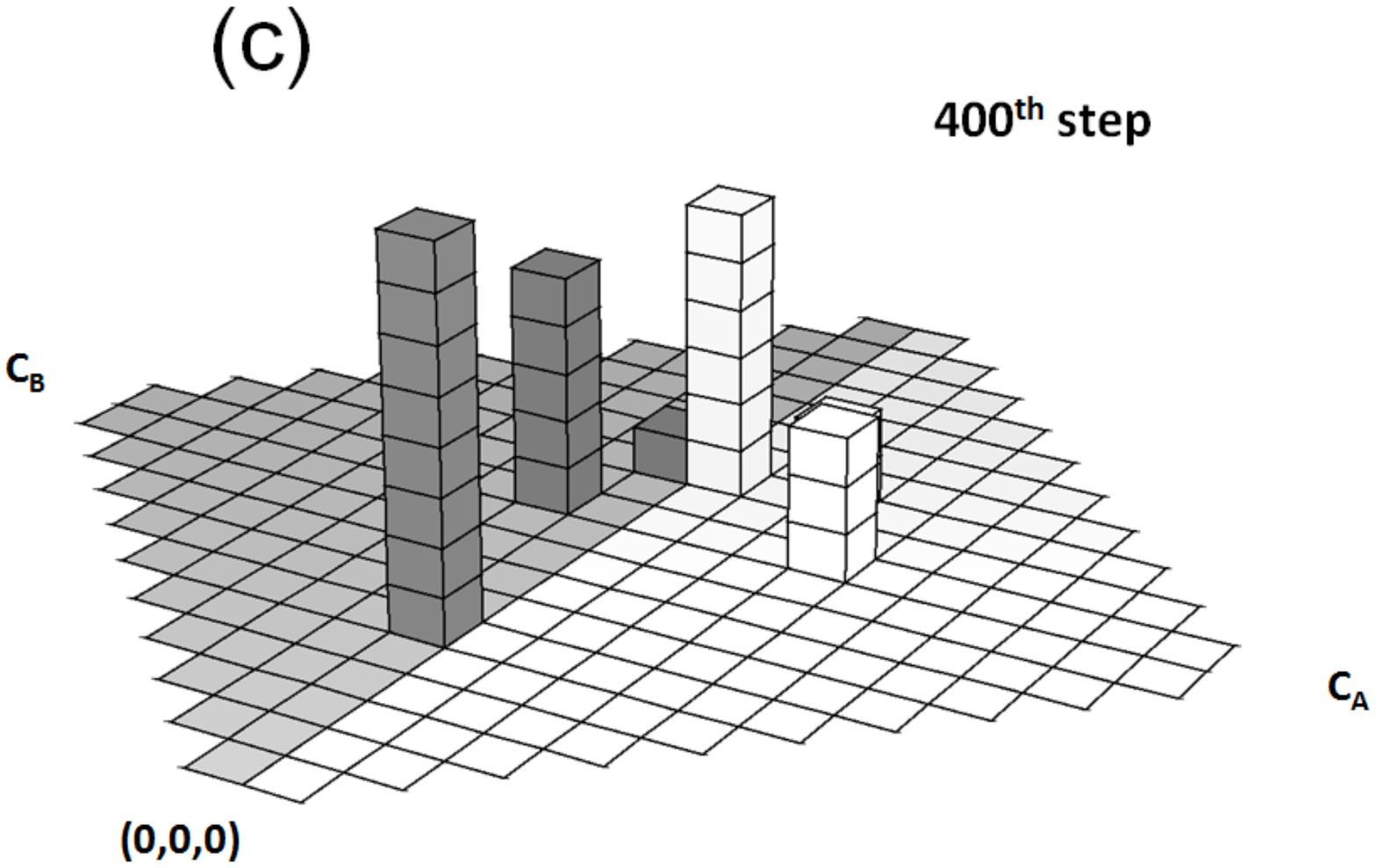}\hspace{0.5cm}
\includegraphics*[width=8cm,height=8cm]{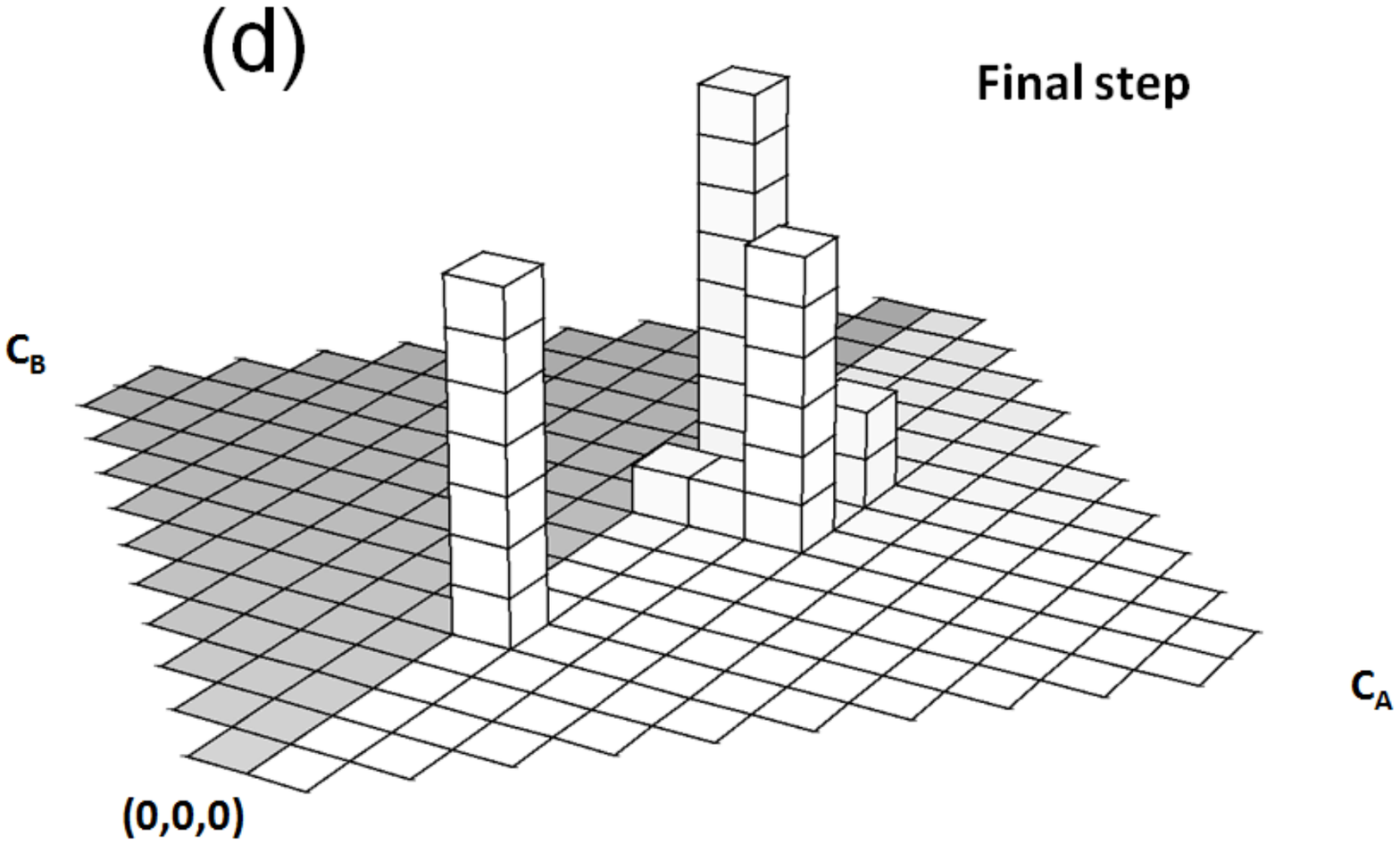}\\
{Fig.\,5: (a) Plot of revealed preference $c_{A,B}$ versus number of steps $T$ 
with $\alpha=0.5$, $\mu=0.5$, (b)  frequency of $c_{A,B}$ at the first step, (c)  frequency of $c_{A,B}$ at the $400^{th}$ step, (d)  frequency of $c_{A,B}$ at the final step. Consensus reached is choice A.}
\end{figure}

\begin{figure} 
 \centering
 \includegraphics*[width=16cm,height=12cm]{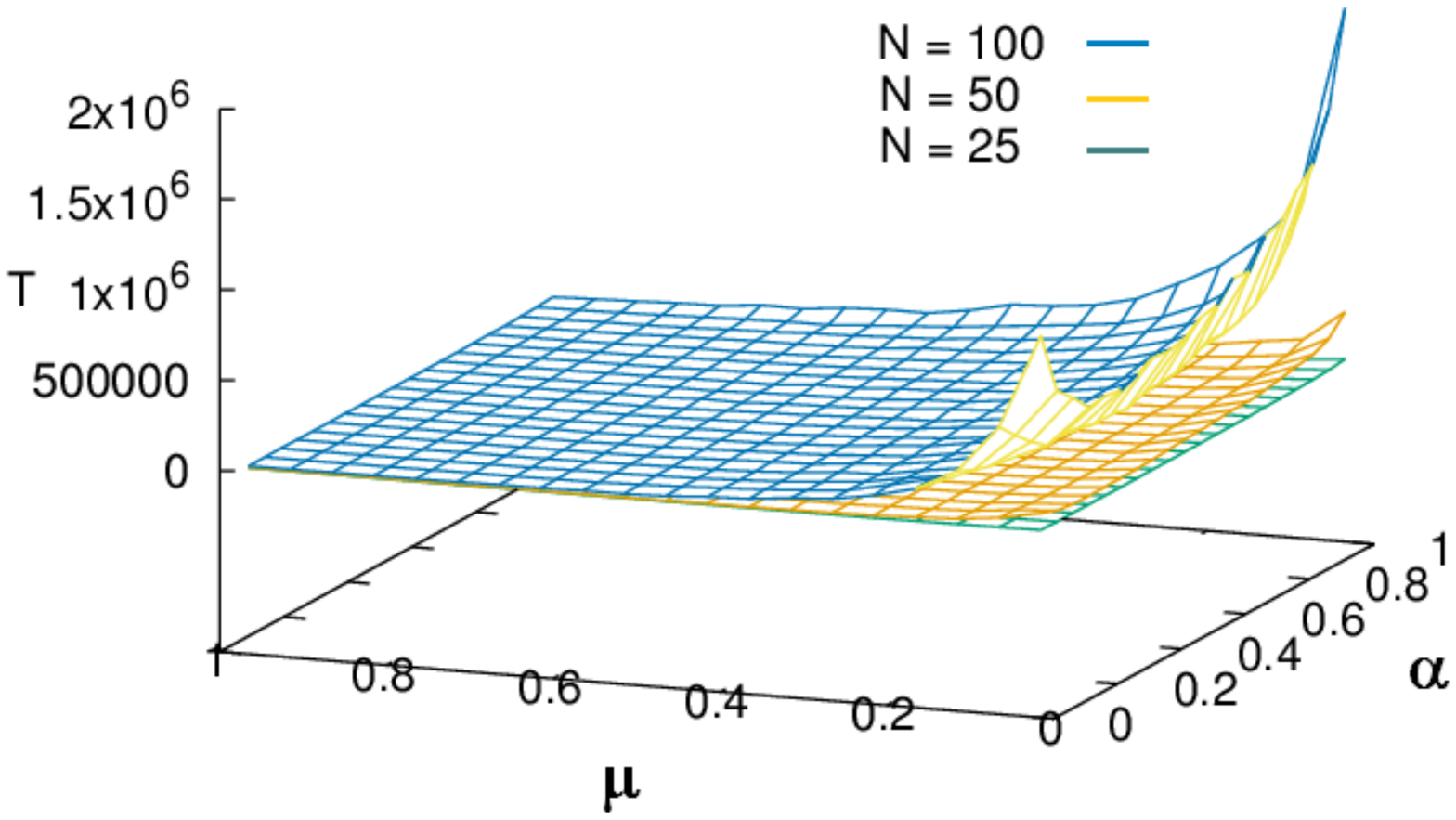}\\
{Fig.\,6: The 3-dimensional plot of $\langle T\rangle$ versus $\alpha$ and $\mu$ for $N=25, 50$ and $100$ agents with an ensemble size of $200$ for each set of parameters.}
\end{figure}

%
%

\subsection{Discussion}

The cases considered in our numerical simulation show that consensus can always be attained in the present model with two possible choices. However, the time needed to achieve consensus, or the so-called convergence time, is longer if the listener is more concerned with the public opinion (as measured by the parameter $\alpha\sim 1$ in our model), or is less likely to be influenced by the lobbyist (as measured by the parameter $\mu\sim 0$ in our model).

The model presented here is bi-directional in that the direction of the propagation of opinion can be forward or backward, depending on wether the listener is convinced or not.
One could also consider a model which is  uni-directional: the listener will try to change his/her next nearest neighbour in the forward direction with his modified/original opinion depending on whether he/she is convinced or not by the previous lobbyist.

A more interesting generalisation of the model is to consider a system with more than two possible choices of opinions. One would expect more complicated phase structures in the opinion space, such as polarisation or fragmentation of opinions. Such systems are now under investigation. 


\section*{Acknowledgment}

The author would like to thank C.-I. Chou for discussions and numeric works at various stages of this work.
 
\newpage
 


\begin{thebibliography}{99}
 
 \bibitem{CH1}
Chou C.-I., and Ho C.-L., Chin.Phys. B 23, 110302 (2014).
 
 \bibitem{CH2}
 Chou C.-I. and Ho C.-L., ``A model of interacting multiple choices of continuous opinions", arXiv:1601.00570 [physics.soc-ph] (2016).
 

\bibitem{Kempe} 
Kempe J., Contemporary Physics 44, 307 (2003).

\bibitem{VA}  
Venegas-Andraca S. E., Q. Info. Proc. 11,1015 (2012).

\bibitem{RNB}
Reitzner D., Nagaj D., Bu\u{z}ek V., Acta Phys Slovaca 61, 603 (2011).

\bibitem{Portugal}
 Portugal R., Quantum Walks and Search Algorithms , Springer, 2013.
 
 \bibitem{MW}
 Manuochehri K., and Wang J.
 Physical Implementation of Quantum Walks, Springer, 2014.

\bibitem{ADZ} 
Aharonov Y., Davidovich L., and Zagury N., Phys. Rev. A  48, 1687 (1993).

\bibitem{NV} 
Nayak A.,  and Vishwanath A.,  Quantum walk on the line, DIMACS Technical Report 2000-43, 
quant-ph/0010117 (2000).

\bibitem{LZG0}
Li M., Zhang Y.-S., and Guo G.-C., Chin. Phys. B 22, 030310 (2013).

\bibitem{XZ}	
Xue P., and  Zhang Y.-S., Chin. Phys. B 22, 070302 (2013).

\bibitem{QX}	
Qin H., and  Xue P.,  Chin. Phys. B 23, 010301 (2014).
	
\bibitem{AAKV} 
Aharonov D., Ambainis A., Kempe J., and Vazirani U.,
Proceedings of the 33th  ACM Symposium on
the Theory of Computing (STOC '01) ACM,  50 (2001).

\bibitem{WZTQW}
Wu J.-J., Zhang B.-D.,  Tang Y.-H., Qiang X.-G., and Wang H.-Q., Chin. Phys. B 22, 050304 (2013).

\bibitem{FG} 
Farhi E.,   and Gutmann S., Phys. Rev. A  58, 915 (1998).

\bibitem{RSLLZG}
Ren C.-N., Shi P., Liu K., Li W.-D., Zhao J. and  Gu Y.-J., Acta Phys. Sin. 62, 090301 (2013).

\bibitem{SKW}  
Shenvi N., Kempe J.,  and Whaley K. B., Phys. Rev. A  67, 052307 (2003).

\bibitem{CCDFGS} 
Childs A. M., Cleve R., Deotto E., Farhi E.,
Gutmann S. and Spielman D. A., Proceedings of the 35th ACM Symposium
on Theory of Computing (STOC '03)  ACM, 59 (2003).

\bibitem{Konno1}
Konno N., Q. Info. Proc. 1, 345 (2002).
\bibitem{Konno2}
Konno N., Q.  Info. and Comp. 2, 578 (2002).
\bibitem{Konno3}
Konno N., J. Math. Soc.  Japan 57, 1179 (2005).


\bibitem{TFMK}
Tregenna B., Flanagan W., Maile R., and Kendon V.,  New J. Phys. 5, 83 (2003).

\bibitem{IKK}
Inui N., Konishi Y., and Konno N.,  Phys. Rev. A 69, 052323 (2004).

\bibitem{IK}
Inui N., and Konno N., Physica A 53, 133 (2005).

\bibitem{IKMS}
Ide Y., Konno N., Machida T., and Segawa E., Quantum Information and
Computation 11, 761 (2011).
 
\bibitem{RMM} 
Ribeiro P, Milman P., and Mosseri R.,Phys. Rev. Lett. 93, 190503 (2004).

\bibitem{S1}
 \u{S}tefa\u{n}\'{a}k M.,  Jex J., and  Kiss T.,  Phys. Rev. Lett.100, 020501 (2008).

\bibitem{S2}
\u{S}tefa\u{n}\'{a}k M., Kiss T., and Jex I.,  Phys. Rev. A. 78, 032306 (2008).

\bibitem{S3}
\u{S}tefa\u{n}\'{a}k M., Kiss T., and Jex I.,  New J. Phys.  11, 043027 (2009).

\bibitem{KS}
Konno N.,  and Segawa E.,  Q.  Info. Comp. 11, 485 (2011).

\bibitem{Konno4}
Konno N., Q. Info.  Proc. 9, 405 (2010).
 
\bibitem{SK1}
Shikano Y., and Katsura H., Phys. Rev. E 82, 031122 (2010)..
\bibitem{SK2}
Shikano Y., and Katsura H.,  AIP Conf. Proc. 1363, 151 (2011).
 
\bibitem{LS}
Linden N., and Sharam J., Phys. Rev. A 80, 052327 (2009).

\bibitem{Wojcik}
W\'ojcik A et al.,  Phys. Rev. A 85, 012329 (2012).

\bibitem{Machida}
Machida T., J. Comput. Theor. Nanosci. 10, 1571 (2013).

\bibitem{FP}
di Franco C., and Paternostro M.,  Phys. Rev. A 91, 012328 (2015).

\bibitem{Shikano1}
Shikano Y.,  AIP Conf. Proc. 1327, 487 (2011).

\bibitem{Shikano2}
Shikano Y.,  J. Comput. Theor. Nanosci. 10, 1558 (2013).

\bibitem{Xue}
Xue P., Zhang R., Qin H., Zhan X., Bian Z.H., Li J., and Sanders B.C.,
Phys. Rev. Lett. 114, 140502 (2015).

\bibitem{CW}
 Cedzich C. and Werner R.F., arXiv:1510.08905 [quant-ph] (2015).
 
\bibitem{LZG}
 Li M., Zhang Y.-S., and  Guo G.-C., Chin. Phys. Lett. 30, 020304 (2013).
 
\bibitem{Eberly}
Eberly J.H., Narozhny N.B., and Sanchez-Mondragon J.J.,  Phys. Rev. Lett. 44, 1323 (1980). 

\bibitem{Konno5}
Konno N., Shimizu Y., and Takei M., 
``Periodicity for the Hadamard walk on cycles", work presented at 
``Workshop of Quantum Simulation and Quantum walks 2015", to appear in this journal (2016).

\bibitem{CFL}
 Castellano C.,  Fortunato S., , and Lereto V., , Rev. Mod. Phys.  81, 591 (2009).
 
 \bibitem{SC}
 Sen P., and Chakrabarti B.K.,
 Socialphysics: An Introduction, Oxford University Press, 2014.
  

\bibitem{G1}
Galam S., and Moscovici S., European J. Social Psychology 21, 49 (1991).

\bibitem{voter}
Clifford P., and A. Sudbury A., Biometrika 60, 581 (1973).

\bibitem{Sznajd}
Sznajd-Weron  K., and Sznajd J., Int. J. Mod. Phys. C 13, 1157 (20000.

\bibitem{G2}
Galam S., Eur. Phys. J. B 25, 403 (2002).

\bibitem{G3}
Gekle, S. , Peliti L., and Galam S., Eur. Phys. J. B 45, 569 (2005). 

\bibitem{G4}
Galam S. , Global Economics and Management Review 18, 11 (2013).

\bibitem{SIT}
Latan\'e B., Am. Psychol. 36, 343 (1981).

\bibitem{D}
Deffuant G., Neau D., Amblard F., and Weisbuch G., Adv. Compl. Sys. 3, 87 (2000).

\bibitem{HK}
Hegselmann R.,  and Krause, U.,  J. Artificial Societies and Social Simulation 5, 2 (2002).

\bibitem{LCCC}
Lallouache M., Chakrabarti A.S., Chakraborti A., and Chakrabarti B.K., Phys. Rev. E 82, 056112 (2010). 

\bibitem{FLPR}
Fortunato, S., Latora V., Pluchino A., and Rapisarda A., Int. J. Mod. Phys. C 16, 1535 (2005).

\bibitem{L}
J. Lorenz, In: Managing Complexity: Insights, Concepts, Applications ; Helbing, Dirk (Ed.); Springer Series "Understanding Complex Systems", arXiv: 0708:3172 [physics.soc-ph] (2008).


\end{thebibliography}
\end{document}